\documentclass[11pt]{article}

\usepackage{graphicx}

\usepackage{cite}

\usepackage{amsthm}

\usepackage{array,color}

\begin{document}

%



%

\title{\Large\bf
Azimuthal  asymmetries for quark pair production in pA collisions }

\author{Emin Akcakaya, Andreas Sch\"{a}fer  and Jian Zhou
 \\[0.3cm]
{\normalsize\it Institut f\"{u}r Theoretische Physik,Universit\"{a}t Regensburg, Regensburg, Germany}}

\maketitle




%



%

\begin{abstract}
\noindent
We study the azimuthal asymmetries for quark pair production in proton-nucleus collisions using a hybrid approach in which
the nucleus is treated in the Color Glass Condensate (CGC) framework while the Lipatov approximation
is applied on the proton side.
Our treatment goes beyond the large $N_c$ limit.
We particularly focus on the  so-called correlation limit where the
imbalance of the transverse momentum of the quark pair is much smaller than the
out-going individual quark transverse momenta.
In this kinematic region, a matching between the hybrid approach and a factorization in terms of transverse momentum dependent
parton distributions (TMDs) has been found.
It is shown which of the various unpolarized and linearly polarized gluon TMD distributions contribute to $\cos 2 \phi$
and $\cos 4 \phi$ modulations of quark pair production.
\end{abstract}

%



%

\section{Introduction}
\noindent
 Attempting to understand the internal structure of the
nucleon and of nuclei in terms of quarks and gluons, the fundamental
degrees of freedom of QCD, has been and still is a very active area
of hadronic/nuclear high energy physics. The information on this
structure is encoded, e.g., in different types of parton
distribution functions which, in the most important twist-2 cases,
can be interpreted as number densities of partons inside a
nucleon/nucleus. The best known objects are the so-called forward
parton distribution functions (PDFs), which, however, provide only a
one-dimensional picture of a hadron, depending merely on the
longitudinal momentum fraction $x$ carried by the partons and the
resolution of the probe (i.e. on $Q^2$). A natural next step in
complexity are transverse momentum dependent parton distributions
(TMDs). They contain information on transverse parton motion, novel
transverse spin correlations etc,  and depend on $x$ and the
transverse parton momentum $k_\perp$ (as well as the relevant $Q^2$
of the probe). Understanding the $Q^2$ evolution of TMDs as well as
controlling higher-order/higher-twist contributions and TMD
factorization in general is technicallly rather
difficult~\cite{Collins:1981uk,Ji:2004wu,
collins_idilbi_neubert,Collins:2012ss}.

At present,
many issues of phenomenological interest are still basically unsettled.
In this paper we concentrate on
gluon TMDs (more precisely linearly polarized gluon TMDs at small $x$)
of which hardly anything is known and we will study their relevant factorization
properties.
As the data from the $p+Au$ runs at LHC are in the process of getting published,
see e.g.~\cite{ATLAS} such studies are of potentially  topical interest.
As we will not address evolution, the subtleties which
made the understanding of TMD factorization difficult for many years
are not relevant for us.
In fact, TMD factorization is still not completely understood, despite the progress
made in recent years. Most importantly, it was shown
in~\cite{Collins:2007nk,Rogers:2010dm} that it
gets violated for di-jet production in hadron-hadron collisions
in higher orders.  So much work still has to be done
to reach a complete and precise theoretical understanding, but
 we are for the time being primarily interested in identifying experimental
signatures and their approximate magnitude.
To do so, we base our study on well-established
models. Comparable model-based studies, e.g.,
in Refs.~\cite{Boer:2011fh,Anselmino:2011ay} (and references therein)
have already generated valuable insights and we try to add to these.

The major obstacle to achieve a TMD factorization
for the processes discussed in~\cite{Rogers:2010dm} is in fact
caused by the longitudinally polarized gluon
attached to the hard scattering part
from both incoming nucleons, which cannot be disentangled and absorbed into the
gauge links appear in the matrix element definition of gluon TMDs.
For the same reason, we will encounter similar problems in the nucleon-nucleus collisions, but
in the small x region, they can be avoided to some extent as we will explain next,
following the arguments of~\cite{Dominguez:2010xd,Dominguez:2011wm}:

If the momentum fraction of the parton becomes sufficiently small the QCD
splitting of partons and their recombination is expected to be in balance.
In such a small-$x$ limit, the nonlinear saturation effects become very crucial
to describe the dynamics  of the hadronic/nuclear systems.
An effective field theory --- the so-called Color Glass Condensate (CGC) approach
has been developed (see, e.g., Ref.~\cite{McLerran:1993ni})
and was widely used to study the saturation physics.
Due to the existence of a semi-hard scale, namely the saturation scale,
the gluon TMD distributions with different gauge link structures can be
perturbativly calculated in the small-$x$ region within the CGC framework.
Moreover, an effective TMD factorization recently has been established at small $x$
for di-jet production in nucleon-nucleus collisions~\cite{Dominguez:2010xd,Dominguez:2011wm}.
As a consequence, the gluon TMDs can be extracted by measuring the transverse momentum
imbalance of di-jet produced in nucleon-nucleus collisions.
Comparing them with those derived in the CGC approach provides a unique chance to test
saturation physics.

To achieve such an effective TMD factorization, one first has to calculate hard
scattering amplitudes in the CGC framework and then
extrapolate the full CGC result to the so-called correlation limit where the
$k_{\perp}$-imbalance of di-jets is much smaller than each of the jet transverse momenta.
The key observation was that in the
correlation limit, two out-going jets stay very close in position
space, such that multiple-point functions appearing in the CGC
formalism collapse into the two point function. The derivative of
the two point function with respect to the transverse coordinate was
related to the various gluon TMDs. Following this line of argument
one would conclude that the basic building blocks are gluon
multiple-point functions, namely Wilson line, rather than gluon
TMDs. Note that the multiple soft gluon interaction between the
nucleon and the active partons is neglected in the CGC calculation
since the background gluon field inside a large nucleus is much
stronger than that inside a nucleon. This leads to the absence of
contributions that cause a violation of generalized TMD
factorization~\cite{Rogers:2010dm}. A comprehensive review covering
among others the relation between the CGC formalism and TMD
factorization can be found in Ref.~\cite{Avsar:2012hj}.

In this paper, following the same spirit,
we study quark pair production in the correlation limit in $pA$ collisions,
with special focus on the polarized case.
Quark pair production in high energy hadron-hadron collisions has been investigated in the framework
of collinear factorization in~\cite{Nason:1987xz,Frixione:1997ma}
and in $k_\perp$ factorization in~\cite{Catani:1990eg,Collins:1991ty,Levin:1991ry}.
Later, a CGC calculation has shown that for quark pair production in nucleon-nucleus collisions,
the result for $k_\perp$ factorization can be recovered from the CGC one in the dilute limit where
the gluon densities are not too high, but that this
fails in a dense medium~\cite{Gelis:2003vh,Blaizot:2004wv}.
In addition to the two point function, three point functions and four point functions  also show up in the
CGC calculation.
These multiple-point functions provide a deep access  into the saturation physics in the
kinematical region beyond the correlation limit.

We will first reproduce the full CGC result using a hybrid approach~\cite{Schafer:2012yx} in which
the nucleus is treated in the Color Glass Condensate(CGC) framework~\cite{McLerran:1993ni,Mueller:2001fv}
while the Lipatov approximation~\cite{Kuraev:1977fs,Gribov:1984tu,Catani:1990eg,Collins:1991ty} is applied
on the proton side.
The second step is to extrapolate this result to the correlation limit by using a power expansion
of the hard coefficients.
The fact that the hard coefficients become independent of the gluon transverse momentum enables us to
integrate out one or two gluon transverse momenta. Correspondingly, the three point functions and four point functions
collapse and can be expressed as derivatives of the two point function. The latter can be related to the various gluon TMDs.
Using the different polarization tensor structures, the gluon TMDs are classified into an unpolarized gluon distribution and
the distribution
of  linearly polarized gluons. The latter one is responsible for $\cos 2 \phi$ and $\cos 4\phi$ azimuthal asymmetries
for quark pair production in proton-nucleus collisions.

The linearly polarized gluon distribution~\cite{Mulders:2000sh,Anselmino:2005sh} ($h_{1}^{\perp g}$
in the notation of Ref.~\cite{Meissner:2007rx})  recently has attracted quite a lot of attention.
It is the only spin dependent gluon TMD for an unpolarized nucleon/nucleus, and may be considered as the
counterpart of the quark Boer-Mulders function~$h_1^{\perp q}(x,k_{\perp})$ \cite{Boer:1997nt}. However, in contrast to the
latter, $h_1^{\perp g}$ is time-reversal even implying that initial/final state interactions are not needed for its
existence~\cite{Brodsky:2002cx,Collins:2002kn}. Despite this fact, it does receive
the contributions from the initial/final state interaction, leading to the process dependent gauge links.
This distribution function is of phenomenological interest, especially for small-$x$
physics at RHIC and LHC because a calculation in the saturation model~\cite{Metz:2011wb} showed that its contributions are
(at small-$x$) as large as those proportional to the unpolarized gluon distribution.
The saturation model calculation also reveals that the linearly polarized gluon distributions with different
gauge link structures differ significantly
at low transverse momentum, though they all recover the normal
perturbative tail at high transverse momentum.

A few ways of accessing $h_{1}^{\perp g}$ have been put forward, namely through measuring azimuthal $\cos 2\phi$ asymmetries in processes such as jet or
quark pair production in electron-nucleon scattering as well as nucleon-nucleon scattering~~\cite{Boer:2009nc,Boer:2010zf}.
Other promising observables are $\cos2\phi$ asymmetries in photon pair production in hadron collisions~\cite{Qiu:2011ai}
and in virtual photon-jet production in nucleon nucleus collisions~\cite{Metz:2011wb}.
Such measurements should be feasible at RHIC,  LHC, and a potential future Electron Ion Collider
(EIC)~\cite{Anselmino:2011ay,Boer:2011fh}.
More recently, it has been found that the linearly polarized gluon distribution may
affect the transverse momentum distribution of Higgs bosons produced
from gluon fusion~\cite{Boer:2011kf,Sun:2011iw,Schafer:2012yx}, color-neutral particles
produced in nucleus-nucleus collisions~\cite{Liou:2012xy}, and heavy quarkonium produced in hadronic collisions\cite{Boer:2012bt}.
The authors of Ref.~\cite{Boer:2011kf} proposed that the effect of linearly
polarized gluons on the Higgs transverse momentum distribution can even be used, in principle, to determine the parity of the Higgs boson experimentally.
Transverse momentum dependent factorization  has been re-examined by taking into account the perturbative
gluon-radiation correction to $h_{1}^{\perp g}$~\cite{Sun:2011iw}.
The complete TMD factorization results for Higgs boson production
are consistent with earlier findings based on the Collins-Soper-Sterman (CSS) formalism~\cite{Catani:2010pd} and
soft-collinear-effective theory~\cite{Mantry:2009qz}.
Also, the transverse momentum resummation formalism applied to di-photon
production in $pp$ collisions~\cite{Nadolsky:2007ba} is closely related.
A recent development\cite{GarciaEchevarria:2011rb}
 has shown that it might also be promising to perform the resummation procedure on the light-cone.

The article is organized as follows. In the next section, we start by reviewing our version
of the hybrid approach which allows us to describe effects caused by the finite gluon transverse
momentum on the proton side.
Then we reproduce the known result for the quark pair production amplitude in $pA$
collisions using this hybrid approach.
The next step is a power expansion in the correlation limit. We show that the resulting differential cross
section depends only on
gluon TMDs rather than higher multiple-point functions.
We particularly focus on the polarized cross section which contains the linearly polarized
gluon distributions.  Our treatment goes beyond the large $N_c$ limit
used in earlier studies.
In section III, we discuss our result in the dilute limit, the forward limit and the large $N_c$ limit.
It is shown that our expressions are consistent with the existing results in these different limits.
In section IV,  we rederive the  cross section in the TMD factorization framework.
As expected, a matching between the CGC formalism and the TMD factorization approach
is found in the correlation limit. The phenomenological implication of our work is briefly discussed in the end of this section.
In the Section V, we summarize our results.

\section{Quark pair production in the CGC framework}

\noindent
Let us first consider the general case of quark pair production,
\begin{eqnarray}
{\cal P }(P_B)+{\cal A}(P_A/{\rm per \, nucleon}) \rightarrow q(l_1)+\bar q(l_2)+X \ .
\end{eqnarray}
We assume that the nucleus is moving with a velocity very close to the speed of light into the positive
$z$ direction, while the proton is moving in the opposite direction.
It is convenient to use light-cone coordinates for which $P_A^\mu=P_A^+p^\mu$ and $P_B^\mu=P_B^- n^\mu$
with $p = (1, 0, 0, 0)$ and $n = (0, 1, 0, 0)$.  The corresponding partonic subprocess is represented by
$g_A(k_1)+g_p(k_2) \rightarrow q(l_1)+\bar q(l_2) $, where $k_1^\mu=x_1P_A^\mu+ k_{1T}^\mu$ denotes the total momentum
carried by multiple gluons from  the nucleus, and
$k_2^\mu=x_2 P_B^\mu+k_{2T}^\mu$ is the momentum of the gluon from the proton.
In the following the notations $k_{1\perp}$ and $ k_{2\perp}$ are used to denote three-dimensional vectors
with  $k_{1\perp}^2=-k_{1T}^2$ and $k_{2\perp}^2=-k_{2T}^2$.
To simplify the calculation, we choose to work in the light-cone gauge of the proton $(A^-=0)$. Correspondingly,
the polarization tensor of  a gluon carrying the momentum $l$ is  given by,
\begin{eqnarray}
\varepsilon^{\mu \nu}(l)= -g^{\mu \nu}+ \frac{p^\mu l^\nu+p^\nu l^\mu}{p \cdot l} \ .
\end{eqnarray}

As mentioned in the introduction, to facilitate our calculation, a hybrid approach~\cite{Schafer:2012yx} has been adopted, in
which the nucleus is treated in the CGC model,
while on the side of the dilute projectile proton one makes the so-called Lipatov
approximation~\cite{Kuraev:1977fs,Gribov:1984tu,Catani:1990eg,Collins:1991ty}.
At small $x$, the gluon radiation cascade shows a strong ordering in rapidity.
In other words, the radiating color source carries a much larger longitudinal momentum than the radiated gluon.
It has been shown that a fast moving color source can be treated as an eikonal line in the strongly rapidity ordered
region. Making such a replacement is referred to as Lipatov approximation~\cite{Collins:1991ty}.

For the process of gluon production in $pA$ collisions, the relevant eikonal line is
the past-pointing one which is built up through initial state interactions between the
color sources inside the proton and the background gluon field.
The interaction between the classical gluon field and the final state gluon emitted from
the color source inside the proton does not change this general statement because
the imaginary part of the scattering amplitude cancels between the different cut diagrams
once the final states are integrated out.
The prescription to treat the eikonal propagator is fixed by this choice.
The relevant Feynman rules, illustrated in Fig.\ 1, were given in Ref.~\cite{Collins:1991ty}.
Note that the prescription for past-pointing eikonal propagators differs from that for
future-pointing eikonal lines.

\begin{figure}[t]
\begin{center}
\includegraphics[width=12cm]{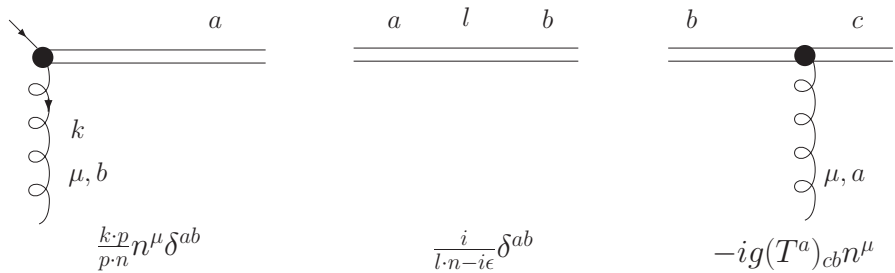}
\end{center}
\vskip -0.4cm \caption{ Feynman rules for the eikonal line, which is  represented by a double line. $a,b$ and $c$ denote color indices.}
\label{efig1}
\end{figure}

The multiple scattering between the outgoing quark pair  and the classical color
field of the nucleus can be readily resummed to all orders~\cite{Balitsky:1995ub,McLerran:1998nk}.
This gives rise to a path-ordered gauge factor along the straight line
that extends in $x^-$ from minus infinity to plus infinity. More precisely, for a quark with incoming momentum
 $l$ and outgoing momentum  $l+k$, the path-ordered gauge factor reads,
\begin{equation}
 2 \pi  \delta(k^-)
 p^\mu [U-1](k_\perp) \,,
\end{equation}
with
\begin{equation}
[U-1](k_\perp)
=\int d^2 x_\perp e^{-  i  k_\perp \cdot  x_\perp} [U(x_\perp)-1] \,,
\end{equation}
and
\begin{equation}
U(x_\perp)= \langle {\cal P} e^{-ig_s \int_{-\infty}^{+\infty} dx^- A^+(x^-, \ x_\perp)} \rangle_A \, ,
\end{equation}
where $A^+=A^+_c t^c$ with $t^c$ being the generators in the fundamental representation.
Similarly, multiple scattering between incoming gluon (or eikonal line)  and classical color
field of the nucleus also can be resummed to all orders,
\begin{equation}
 2 \pi  \delta(k^-)
 p^\mu [\tilde U-1](k_\perp) \,,
\end{equation}
with
\begin{equation}
[\tilde U-1](k_\perp)
=\int d^2 x_\perp e^{-  i  k_\perp \cdot  x_\perp} [\tilde U(x_\perp)-1] \,,
\end{equation}
and
\begin{equation}
\tilde U(x_\perp)= \langle {\cal P} e^{-ig_s \int_{-\infty}^{+\infty} dx^- \tilde A^+(x^-, \ x_\perp)} \rangle_A \, ,
\end{equation}
where $(\tilde A^+)_{ab}=A^+_c (-if^{abc})$ with $f^{abc}$ being the generators in the adjoint representation.

We use these as the building blocks to compute the amplitude for quark pair production in high energy $pA$ collisions.
It is straightforward to obtain the production amplitude for diagram(a) illustrated in the Fig.2,
\begin{eqnarray}
{\cal M}^{(a)}= -ig_s\bar u(l_1) \gamma^\rho t^a  \frac{l\!\!\!/_2-k\!\!\!/_1-m}{(l_2-k_1)^2-m^2+i \epsilon} p\!\!\!/[U^\dag-1](k_{1\perp}) v(l_2)
\frac{ \varepsilon_{\rho \sigma}(k_2)}{k^2_2+i \epsilon}n^{\sigma} \frac{ k_2 \! \cdot \! p}{p \! \cdot \! n}
 \phi_p(x_2,k_{2\perp}) \, ,
\end{eqnarray}
where the factor $2\pi \delta(k_1^-)$ is suppressed. $m$ is the quark mass.
$k_2=l_1+l_2-k_1$ denotes the momentum of the gluon from the proton with
$l_1$ and $l_2$ being the quark and anti-quark momentum, respectively.
$\phi_p(x_2,k_{2\perp})$ represents the probability amplitude  for finding a gluon carrying a
certain momentum inside the proton, with $x_2g(x_2,k_{2\perp})=\phi_p\phi_p^*$.
The other diagrams shown in Fig.2 give,
\begin{eqnarray}
{\cal M}^{(b)}\!\! &=& \!\! -ig_s\bar u(l_1)p\!\!\!/[U-1](k_{1\perp}) \frac{l\!\!\!/_1-k\!\!\!/_1+m}{(l_1-k_1)^2-m^2+i \epsilon}  \gamma^\rho t^a v(l_2)
\frac{ \varepsilon_{\rho \sigma}(k_2)}{k^2_2+i \epsilon}n^{\sigma} \frac{ k_2 \! \cdot \! p}{p \! \cdot \! n}
 \phi_p(x_2,k_{2\perp})
\\
{\cal M}^{(c)}\!\! &=& \!\! g_s \!\! \int \! \frac{d^4 k_1'}{(2 \pi )^3} \delta(k_1^{'-})
\bar u(l_1)p\!\!\!/ [U-1](k_{1\perp}-k_{1\perp}') \frac{l\!\!\!/_1-k\!\!\!/_1+k\!\!\!/_1'+m}{(l_1-k_1+k_1')^2-m^2+i \epsilon}  \gamma^\rho t^a
\nonumber\\&& \ \ \ \ \ \ \ \ \ \ \ \ \ \times
\frac{l\!\!\!/_2-k\!\!\!/_1'-m}{(l_2-k_1')^2-m^2+i \epsilon} p\!\!\!/ [U^\dag-1](k_{1\perp}') v(l_2)
\frac{ \varepsilon_{\rho \sigma}(k_2)}{k^2_2+i \epsilon}n^{\sigma} \frac{ k_2 \! \cdot \! p}{p \! \cdot \! n}
 \phi_p(x_2,k_{2\perp})
\\
{\cal M}^{(d)}\!\! &=& \!\! -ig_s\bar u(l_1) \gamma^\rho t^b  v(l_2)\frac{1}{k_1 \! \cdot \! n - i\epsilon}
\frac{ \varepsilon_{\rho \sigma}(k_2+k_1)}{(k_2+k_1)^2+i \epsilon}n^{\sigma} \frac{ (k_2+k_1) \! \cdot \! p}{p \! \cdot \! n}
[\tilde U-1]_{ba}(k_{1\perp}) \phi_p(x_2,k_{2\perp})
\\
{\cal M}^{(e)}\!\! &=& \!\! -ig_s\bar u(l_1) \gamma^\rho t^b  v(l_2)
\frac{ \varepsilon_{\rho \rho'}(k_2+k_1)}{(k_2+k_1)^2+i \epsilon}\frac{ \varepsilon_{\sigma \sigma'}(k_2)}{k_2^2+i \epsilon}
 p^{\mu} \Lambda^{\mu \rho' \sigma'}n^{\sigma} \frac{k_2 \! \cdot \! p}{p \! \cdot \! n}
[\tilde U-1]_{ba}(k_{1\perp}) \phi_p(x_2,k_{2\perp}) \ .
\end{eqnarray}

\begin{figure}[t]
\begin{center}
\includegraphics[width=13cm]{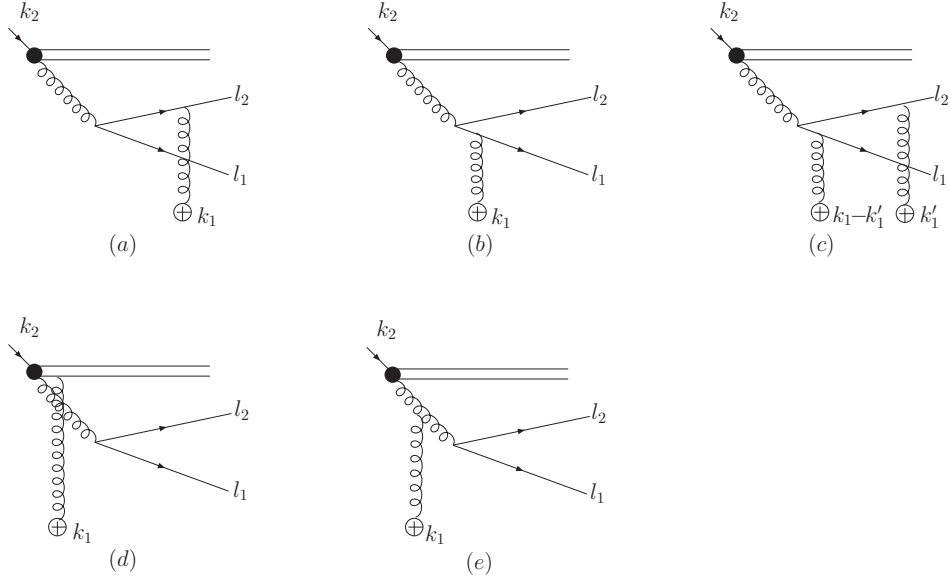}
\end{center}
\vskip -0.4cm \caption{ The diagrams contributing to quark pair production.
The gluon line terminated by a $\oplus$ denotes a classical field insertion.
The contributions from all other diagrams disappear because the multiple poles are located in the same half plane.}
\label{efig2}
\end{figure}

Putting all these terms together, we obtain the following expression for the complete production amplitude,
\begin{eqnarray}
{\cal M}&=&{\cal M}^{(a)}+{\cal M}^{(b)}+{\cal M}^{(c)}+{\cal M}^{(d)}+{\cal M}^{(e)}
\nonumber\\&=&
 \!\! g_s \!\! \int \! \frac{d^4 k_1'}{(2 \pi )^3}  \left \{ \delta(k_1^{'-})
\bar u(l_1)p\!\!\!/  \frac{l\!\!\!/_1-k\!\!\!/_1+k\!\!\!/_1'+m}{(l_1-k_1+k_1')^2-m^2+i \epsilon}  \gamma^\rho
\frac{l\!\!\!/_2-k\!\!\!/_1'-m}{(l_2-k_1')^2-m^2+i \epsilon} p\!\!\!/ \right .\
\nonumber\\ &&   \!\!\! \left .\ \times
 \left [ U(k_{1\perp}-k_{1\perp}')  t^a U^\dag(k_{1\perp}') \right ] v(l_2)
\frac{ \varepsilon_{\rho \sigma}(k_2)}{k^2_2+i \epsilon}n^{\sigma} \frac{ k_2 \! \cdot \! p}{p \! \cdot \! n} \phi_p(x_2,k_{2\perp}) \right \}
-ig_s\bar u(l_1) \gamma^\rho t^b  v(l_2) \tilde U_{ba}(k_{1\perp})
\nonumber \\
&& \times
\left [ \frac{ \varepsilon_{\rho \rho'}(k_2+k_1)}{(k_2+k_1)^2+i \epsilon}\frac{ \varepsilon_{\sigma \sigma'}(k_2)}{k_2^2+i \epsilon} p^{\mu} \Lambda^{\mu \rho' \sigma'}
 + \frac{1}{k_1 \! \cdot \! n - i\epsilon} \frac{ \varepsilon_{\rho \sigma}(k_2+k_1)}{(k_2+k_1)^2+i \epsilon} \right ]
n^{\sigma} \frac{ k_2 \! \cdot \! p}{p \! \cdot \! n}
\phi_p(x_2,k_{2\perp}) \ .
\end{eqnarray}
To arrive the above formula, we have made use of the Dirac equation of motion obeyed by the free spinors and the following identity for the Wilson lines in the different
representations:
\begin{eqnarray}
U(x_\perp)t^a U^\dag(x_\perp)=t^b \tilde U^{ba}(x_\perp) \ .
\end{eqnarray}
After some algebra, this production amplitude can be further rewritten in a more conventional form,
\begin{eqnarray}
{\cal M}&=& \!\! ig_s  \frac{ k_2 \! \cdot \! p}{k^2_2} \phi_p(x_2,k_{2\perp})
\nonumber \\ && \times
\bar u(l_1)  \left \{T_{g}(k_{1\perp}) t^b  \tilde U_{ba}(k_{1\perp})+
\! \int \! \frac{d^2 k_{1\perp}'}{(2 \pi )^2} \left [ T_{q\bar q}(k_{1\perp},k_{1\perp}')
  U(k_{1\perp}-k_{1\perp}')  t^a U^\dag(k_{1\perp}') \right ]  \right \}  v(l_2)
\end{eqnarray}
where,
\begin{eqnarray}
 T_{q\bar q}(k_{2\perp},k_{1\perp},k_{1\perp}') \!\!\! &=& \!\!\! i \int \frac{dk^{'-}dk^{'+}}{2 \pi}
\delta(k_1^{'-})
p\!\!\!/  \frac{l\!\!\!/_1-k\!\!\!/_1+k\!\!\!/_1'+m}{(l_1-k_1+k_1')^2-m^2+i \epsilon}  n\!\!\!/
\frac{l\!\!\!/_2-k\!\!\!/_1'-m}{(l_2-k_1')^2-m^2+i \epsilon} p\!\!\!/
 \\
T_{g}(k_{2\perp},k_{1\perp}) \!\!\! &=& \!\!\! \frac{1}{(k_1+k_2)^2} \left [
\frac{k_{1\perp}^2-(k_{1\perp}+k_{2\perp})^2}{k_2 \cdot p } p\!\!\!/ -\frac{ k_{2\perp}^2}{k_1 \cdot n}n\!\!\!/+2 k\!\!\!/_{2\perp}
\right ]-\frac{1}{k_2 \cdot p} p\!\!\!/  \ .
\end{eqnarray}
This result is in agreement with the production amplitude obtained in Ref.~\cite{Blaizot:2004wv} up to a trivial prefactor.
Note that $T_{q\bar q}(k_{2\perp}=0,k_{1\perp},k_{1\perp}')+T_{g}(k_{2\perp}=0,k_{1\perp})=0$.

Squaring the amplitude, we obtain the following expression for the pair production cross section,
\begin{eqnarray}
\frac{d \sigma}{d{\cal P.S.}}&=& \frac{\alpha_s \pi}{N_c^2-1}
\int\frac{2 d^2 k_{1\perp}}{(2\pi)^3} d^2k_{2\perp} \frac{ d^2k_{1\perp}' d^2k_{1\perp}''}{(2\pi)^4}
\frac{1}{(2\pi)^2} \delta^2(k_{1\perp}+k_{2\perp}-q_\perp) x_2 g (x_2,k_{2\perp} )
\nonumber \\ & \times & \!\!\!
\int d^2x_\perp d^2y_\perp d^2x_\perp' d^2y_\perp'
e^{-i   x_\perp  \cdot   ( k_{1\perp}- k_{1\perp}')} e^{-i  y_\perp  \cdot  k_{1\perp}'}
e^{i   x_\perp'  \cdot  ( k_{1\perp}- k_{1\perp}'')} e^{i   y_\perp'  \cdot  k_{1\perp}''} \frac{1}{k_{2\perp}^2}
\nonumber \\ &&  \!\!\! \times  \left \{
{\rm Tr} \left [(l\!\!\!/_1+m)T_{q\bar q}(l\!\!\!/_2-m)\gamma^0 T_{q\bar q}^{\dag '}\gamma^0 \right ] C(x_\perp,y_\perp,y_\perp',x_\perp')
\right .\
\nonumber \\ && \left .\  +
{\rm Tr} \left [(l\!\!\!/_1+m)T_{q\bar q}(l\!\!\!/_2-m)\gamma^0 T_{g}^{\dag '}\gamma^0  \right  ] C(x_\perp,y_\perp,y_\perp',y_\perp')
\right .\
\nonumber \\ && \left .\  +
{\rm Tr} \left [(l\!\!\!/_1+m)T_{g}(l\!\!\!/_2-m)\gamma^0 T_{q\bar q}^{\dag '}\gamma^0 \right ] C(x_\perp,x_\perp,y_\perp',x_\perp')
\right .\
\nonumber \\ && \left .\  +
 {\rm Tr} \left [(l\!\!\!/_1+m)T_{g}(l\!\!\!/_2-m)\gamma^0 T_{g}^{\dag '}\gamma^0 \right ] C(x_\perp,x_\perp,y_\perp',y_\perp')
\right \} \ .
\end{eqnarray}
where $g(x_2,   k_{2\perp})$  denotes the un-integrated gluon distribution of a proton.
In the phase space factor $d{\cal P.S.}=d^2l_{1\perp} d^2l_{2\perp} dy_1 dy_2$, the quantities
$y_1, y_2$ are the rapidities of the produced quark and anti-quark respectively.
The quark pair imbalance is defined as $  q_\perp= l_{1\perp}+l_{2\perp}$.
The factor $2 /(2\pi)^3 $ associated with the phase space
integration is chosen such that for a single gluon target,
$\int \frac{2 d^2 k_{1\perp}}{(2\pi)^3} \frac{k_{1\perp}^2}{g^2 N_c}\langle \tilde U (k_{1\perp})  \tilde U^\dag (k_{1\perp}) \rangle_{\rm gluon}=x_1 \delta(1-x_1)$
at lowest non-trivial order (see, for example, Ref.~\cite{Iancu:2002xk}).
To obtain the above result, we  have defined the normalization factor and the flux factor to be
$ k_{2\perp}^2/(2 k_2 \cdot p (N_c^2-1))$ and $1/(2 k_2 \cdot p)$, respectively,
rather than $k_{1\perp}^2 k_{2\perp}^2/(4 x_1 x_2 P_A \cdot P_B  (N_c^2-1)^2)$, $1/(4x_1 x_2 P_A \cdot P_B)$
as used in Ref.~\cite{Collins:1991ty}, since the Lipatov approximation is only applied on the proton side.
We have omitted the arguments of $T_{q \bar q}, T_g$. And $T_{q \bar q}',T_g'$ denote the same quantities with $k_{1\perp}'$ replaced by $k_{1\perp}''$.
The four point function $C(x_\perp,y_\perp,y_\perp',x_\perp')$ is defined as,
\begin{eqnarray}
C(x_\perp,y_\perp,y_\perp',x_\perp')={\rm Tr_c} \left \langle U(x_\perp) t^a U^\dag(y_\perp) U(y_\perp') t^a U^\dag(x_\perp')  \right  \rangle_{x_1} \,
\end{eqnarray}
Here ${\rm Tr_c}$ is a trace over the color indices. The longitudinal momentum fraction of proton and nucleus carried by the incoming gluons
are constrained by the kinematics,
\begin{eqnarray}
x_1=\frac{|l_{1\perp}|e^{-y_1}+|l_{2\perp}|e^{-y_2}}{\sqrt s} \ \ \  ,
\ \ x_2=\frac{|l_{1\perp}|e^{y_1}+|l_{2\perp}|e^{y_2}}{\sqrt s} \ ,
\end{eqnarray}
where $\sqrt s$ is the center of mass energy.

With this derived full CGC result, we proceed to the correlation limit where $|P_\perp| \equiv |l_{1\perp}-l_{2\perp}|/2  \gg |q_\perp|/2$.
In this kinematical region, we may systemically neglect the terms suppressed by  powers of $|k_{2\perp}|/|P_\perp|, |k_{1\perp}|/|P_\perp|$,
$ |k_{1\perp}'|/|P_\perp|$ and $|k_{1\perp}''|/|P_\perp|$
in the four hard coefficients. We first perform a Taylor expansion of the hard coefficients in terms of $k_{2\perp}$. By dropping all terms suppressed
by powers of $|k_{2\perp}|/|P_\perp|$, one ends up with,
\begin{eqnarray}
\frac{d \sigma}{d{\cal P.S.}}&\approx& \frac{\alpha_s }{N_c^2-1}
\int\frac{ d^2 k_{1\perp}}{(2\pi)^4} d^2 k_{2\perp} \frac{ d^2k_{1\perp}' d^2k_{1\perp}''}{(2\pi)^4}
 \delta^2(k_{1\perp}+k_{2\perp}-q_\perp) x_2 g (x_2,k_{2\perp} )
\nonumber \\ & \times & \!\!\!
\int d^2x_\perp d^2y_\perp d^2x_\perp' d^2y_\perp'
e^{-i  x_\perp  \cdot  (k_{1\perp}-k_{1\perp}')} e^{-i  y_\perp  \cdot  k_{1\perp}'}
e^{i  x_\perp'  \cdot  (k_{1\perp}-k_{1\perp}'')} e^{i  y_\perp'  \cdot  k_{1\perp}''}
\nonumber \\ &&  \!\!\! \times  \left \{
{\rm Tr} \left [(l\!\!\!/_1+m) \bar T_{q\bar q}(l\!\!\!/_2-m)\gamma^0 \bar T_{q\bar q}^{\dag '}\gamma^0 \right ]_{k_{2\perp}=0} C(x_\perp,y_\perp,y_\perp',x_\perp')
\right .\
\nonumber \\ && \left .\  +
{\rm Tr} \left [(l\!\!\!/_1+m) \bar T_{q\bar q}(l\!\!\!/_2-m)\gamma^0 \bar T_{g}^{\dag '}\gamma^0  \right  ]_{k_{2\perp}=0} C(x_\perp,y_\perp,y_\perp',y_\perp')
\right .\
\nonumber \\ && \left .\  +
{\rm Tr} \left [(l\!\!\!/_1+m) \bar T_{g}(l\!\!\!/_2-m)\gamma^0 \bar T_{q\bar q}^{\dag '}\gamma^0 \right ]_{k_{2\perp}=0} C(x_\perp,x_\perp,y_\perp',x_\perp')
\right .\
\nonumber \\ && \left .\  +
 {\rm Tr} \left [(l\!\!\!/_1+m) \bar T_{g}(l\!\!\!/_2-m)\gamma^0 \bar T_{g}^{\dag '}\gamma^0 \right ]_{k_{2\perp}=0} C(x_\perp,x_\perp,y_\perp',y_\perp')
\right \}  \ ,
\end{eqnarray}
and with $ \bar T_{q\bar q}, \bar T_{g}$ given by,
\begin{eqnarray}
 \bar T_{q\bar q}(k_{1\perp},k_{1\perp}') \!\!\! &=& \!\!\! i \int \frac{dk^{'-}dk^{'+}}{2 \pi}
\delta(k_1^{'-})
p\!\!\!/  \frac{l\!\!\!/_1-k\!\!\!/_1+k\!\!\!/_1'+m}{(l_1-k_1+k_1')^2-m^2+i \epsilon}  \frac{\hat k\!\!\!/_{2\perp}}{ (k_2 \cdot p)}
\frac{l\!\!\!/_2-k\!\!\!/_1'-m}{(l_2-k_1')^2-m^2+i \epsilon} p\!\!\!/
 \\
\bar T_{g}(k_{1\perp}) \!\!\! &=& \!\!\! \frac{1}{(k_1+x_2P_B)^2} \left [
\frac{-2k_{1\perp} \cdot \hat k_{2\perp}}{k_2 \cdot p } p\!\!\!/ +2\hat k\!\!\!/_{2\perp} \right ] \ ,
\end{eqnarray}
where $\hat k_{2\perp}= k_{2\perp}/ |k_{2\perp}|$ is a unit vector.

Now let's move on to discuss the power expansion on the nucleus side.
The fact that the integrations over $k_{1\perp}',k_{1\perp}'' $ are dominated by the kinematical region
$|k_{1\perp}'|\sim |k_{1\perp}''| \sim Q_s$ --- because the typical small $x$ gluon transverse momentum is
characterized by the saturation momentum --- allows us to employ the power expansion
in the correlation limit $Q_s\sim |k_{1\perp}+k_{2\perp}|/2 \ll |P_\perp|$. To facilitate the power expansion,
we replace $\bar T_{q\bar q}(k_{1\perp},k_{1\perp}')$, $\bar T_{g}(k_{1\perp})$ with the following two expressions with the help of Ward identities
(gauge invariance violation terms in the amplitude are proportional to the gluon off-shellness $\sim k_{1\perp}^2$
 , and thus can be neglected in the correlation limit.),
\begin{eqnarray}
 && \!\!\!\!\!\!\!\!\!\!\!\!\!\!\!\!\!\!\!\!
 \bar T_{q\bar q}(k_{1\perp},k_{1\perp}')   \Rightarrow   \bar T_{q\bar q}(k_{1\perp},k_{1\perp}')
\nonumber \\&& \!\!\!\!\!\!
- i \int \frac{dk^{'-}dk^{'+}}{2 \pi} \delta(k_1^{'-})
\frac{k\!\!\!/_1-k\!\!\!/_1'}{(x_1-x_1')P_A^+} \frac{l\!\!\!/_1-k\!\!\!/_1+k\!\!\!/_1'+m}{(l_1-k_1+k_1')^2-m^2+i \epsilon}
 \frac{\hat k\!\!\!/_{2\perp}}{ (k_2 \cdot p)} \frac{l\!\!\!/_2-k\!\!\!/_1'-m}{(l_2-k_1')^2-m^2+i \epsilon} \frac{k\!\!\!/_1'}{x_1'P_A^+}
 \nonumber \\&\Rightarrow&
-\left [ \frac{\gamma_i}{x_1P_A^+} \frac{l\!\!\!/_1-k\!\!\!/_1+m}{(l_1-k_1)^2-m^2+i \epsilon} \frac{\hat k\!\!\!/_{2\perp}}{ (k_2 \cdot p)} \right ]_{k_{1\perp}=0} (k_{1\perp}^i-k_{1\perp}'^i)
 \nonumber \\ &&
-\left [\frac{\hat k\!\!\!/_{2\perp}}{ (k_2 \cdot p)} \frac{l\!\!\!/_2-k\!\!\!/_1-m}{(l_2-k_1)^2-m^2+i \epsilon} \frac{\gamma_i}{x_1P_A^+} \right ]_{k_{1\perp}=0}
k_{1\perp}'^i+O(\frac{k_{1\perp}^2}{P_\perp^2})\approx  \tilde T_{q\bar q,i}^{A}(k_{1\perp}^i-k_{1\perp}'^i)+\tilde T_{q\bar q,i}^{B}k_{1\perp}'^i
 \\ &&  \!\!\!\!\!\!\!\!\!\!\!\!\!\!\!\!\!\!\!\!
\bar T_{g}(k_{1\perp})  \Rightarrow  \bar T_{g}(k_{1\perp})-
\frac{1}{(k_1+x_2P_B)^2} \left (2+\frac{k_{1\perp}^2}{x_1x_2P_A^+P_B^-} \right ) \hat k\!\!\!/_{2\perp}
 \nonumber \\&& \approx
 \left [ \frac{1}{(k_1+x_2P_B)^2} \frac{-2\hat k_{2\perp,i}}{k_2 \cdot p } p\!\!\!/ \right ]_{k_{1\perp}=0} k_{1\perp}^i
 = \tilde T_{g,i} k_{1\perp}^i  \ ,
\end{eqnarray}
with $i$ denoting the transverse index. By making the above replacement, the differential cross section can be rewritten in the form,
\begin{eqnarray}
\frac{d \sigma}{d{\cal P.S.}}&\approx& \frac{\alpha_s }{(N_c^2-1)}\int\frac{ d^2 k_{1\perp}d^2 k_{2\perp} }{(2\pi)^4}
 \delta^2(k_{1\perp}+k_{2\perp}-q_\perp) x_2 g (x_2,k_{2\perp} )
\int d^2x_\perp d^2x_\perp' e^{-i k_{1\perp} \cdot ( x_\perp-x_\perp') }
\nonumber \\ &&  \!\!\! \times  \left \{
{\rm Tr} \left [(l\!\!\!/_1+m) \tilde T_{q\bar q,i}^A(l\!\!\!/_2-m)\gamma^0 \tilde T_{q\bar q,j}^{A \dag '}\gamma^0 \right ]_{k_{2\perp},k_{1\perp}=0}
\left [ \frac{ \partial^2 C(x_\perp,y_\perp,y_\perp',x_\perp')}{\partial x_\perp^i \partial x_\perp'^j} \right ]_{x_\perp=y_\perp, x_\perp'=y_\perp'}
\right .\
\nonumber \\ &&  \left .\ +
{\rm Tr} \left [(l\!\!\!/_1+m) \tilde T_{q\bar q,i}^A(l\!\!\!/_2-m)\gamma^0 \tilde T_{q\bar q,j}^{B \dag '}\gamma^0 \right ]_{k_{2\perp},k_{1\perp}=0}
\left [\frac{ \partial^2 C(x_\perp,y_\perp,y_\perp',x_\perp')}{\partial x_\perp^i \partial y_\perp'^j} \right ]_{x_\perp=y_\perp, x_\perp'=y_\perp'}
\right .\
\nonumber \\ &&  \left .\  +
{\rm Tr} \left [(l\!\!\!/_1+m) \tilde T_{q\bar q,i}^B(l\!\!\!/_2-m)\gamma^0 \tilde T_{q\bar q,j}^{A \dag '}\gamma^0 \right ]_{k_{2\perp},k_{1\perp}=0}
\left [ \frac{ \partial^2 C(x_\perp,y_\perp,y_\perp',x_\perp')}{\partial y_\perp^i \partial x_\perp'^j} \right ]_{x_\perp=y_\perp, x_\perp'=y_\perp'}
\right .\
\nonumber \\ &&  \left .\  +
{\rm Tr} \left [(l\!\!\!/_1+m) \tilde T_{q\bar q,i}^B(l\!\!\!/_2-m)\gamma^0 \tilde T_{q\bar q,j}^{B \dag '}\gamma^0 \right ]_{k_{2\perp},k_{1\perp}=0}
\left [ \frac{ \partial^2 C(x_\perp,y_\perp,y_\perp',x_\perp')}{\partial y_\perp^i \partial y_\perp'^j} \right ]_{x_\perp=y_\perp, x_\perp'=y_\perp'}
\right .\
\nonumber \\ && \left .\  +
{\rm Tr} \left [(l\!\!\!/_1+m) \tilde T_{q\bar q,i}^A(l\!\!\!/_2-m)\gamma^0 \tilde T_{g,j}^{\dag '}\gamma^0  \right  ]_{k_{2\perp},k_{1\perp}=0}
\left [ \frac{ \partial^2 C(x_\perp,y_\perp,x_\perp',x_\perp')}{\partial x_\perp^i   \partial x_\perp'^j} \right ]_{x_\perp=y_\perp}
\right .\
\nonumber \\ && \left .\  +
{\rm Tr} \left [(l\!\!\!/_1+m) \tilde T_{q\bar q,i}^B(l\!\!\!/_2-m)\gamma^0 \tilde T_{g,j}^{\dag '}\gamma^0  \right  ]_{k_{2\perp},k_{1\perp}=0}
\left [ \frac{ \partial^2 C(x_\perp,y_\perp,x_\perp',x_\perp')}{\partial y_\perp^i   \partial x_\perp'^j} \right ]_{x_\perp=y_\perp}
\right .\
\nonumber \\ && \left .\  +
{\rm Tr} \left [(l\!\!\!/_1+m) \tilde T_{g,i}(l\!\!\!/_2-m)\gamma^0 \tilde T_{q\bar q,j}^{A\dag '}\gamma^0 \right ]_{k_{2\perp},k_{1\perp}=0}
\left [ \frac{ \partial^2 C(x_\perp,x_\perp,y_\perp',x_\perp')}{\partial x_\perp^i   \partial x_\perp'^j} \right ]_{ x_\perp'=y_\perp'}
\right .\
\nonumber \\ && \left .\  +
{\rm Tr} \left [(l\!\!\!/_1+m) \tilde T_{g,i}(l\!\!\!/_2-m)\gamma^0 \tilde T_{q\bar q,j}^{B\dag '}\gamma^0 \right ]_{k_{2\perp},k_{1\perp}=0}
\left [ \frac{ \partial^2 C(x_\perp,x_\perp,y_\perp',x_\perp')}{\partial x_\perp^i   \partial y_\perp'^j} \right ]_{ x_\perp'=y_\perp'}
\right .\
\nonumber \\ && \left .\  +
 {\rm Tr} \left [(l\!\!\!/_1+m) \tilde T_{g,i}(l\!\!\!/_2-m)\gamma^0 \tilde T_{g,j}^{\dag '}\gamma^0 \right ]_{k_{2\perp},k_{1\perp}=0}
\left [ \frac{ \partial^2  C(x_\perp,x_\perp,x_\perp',x_\perp')}{\partial x_\perp^i   \partial x_\perp'^j} \right ]
\right \}  \ ,
\end{eqnarray}
where the transverse momenta $k_{1\perp}'$ and $k_{1\perp}''$ have been integrated out. As a result, the four point functions collapse into the two point functions.
The calculation of the Dirac traces in the above formula is rather easy,
while the evaluation of the soft part in the McLerran-Venugopalan(MV) model is a bit more involved.
In general, the tensor structure of the soft part can be decomposed in the following way,
\begin{eqnarray}
&& \int \left [ \frac{ \partial^2 C(x_\perp,y_\perp,y_\perp',x_\perp')}{\partial x_{\perp,i} \partial x_{\perp,j}' } \right ]_{x_\perp=y_\perp, x_\perp'=y_\perp'}
=
\frac{\delta_{\perp}^{ij}}{2} \, F_1(x_1, k_{1\perp}) +
\bigg(\hat k_{1\perp}^i \hat k_{1\perp}^j - \frac{1}{2} \delta_{\perp}^{ij} \bigg)
H_1(x_1, k_{1\perp}) \,.
\end{eqnarray}
where $\int$ denotes $\int d^2x_\perp d^2x_\perp' e^{-i k_{1\perp} \cdot ( x_\perp-x_\perp') }$.
$\hat k_{1\perp}^i$ is a unit vector $\hat k_{1\perp}^i \equiv k_{1\perp}^i/| k_{1\perp}| $, and $\delta_{\perp}^{ij}=-g^{ij}+(p^in^j+p^jn^i)/p \cdot n$.
The four point function  $C(x_\perp,y_\perp,y_\perp',x_\perp')$ has been evaluated in the MV model in Ref.~\cite{Blaizot:2004wv}.
With the derived four point function, the coefficients $F_i$ and $H_i$ can be computed in a tedious but straightforward way. One finds,
\begin{eqnarray}
F_1&= &\int \left [ \frac{ \partial^2 C(x_\perp,y_\perp,y_\perp',x_\perp')}{\partial x_{\perp}^i \partial x_{\perp}'^j } \right ]_{x_\perp=y_\perp, x_\perp'=y_\perp'}
\!\!\!\!\!\!\! \delta_{\perp}^{ij}
=\int \left [ \frac{ \partial^2 C(x_\perp,y_\perp,y_\perp',x_\perp')}{\partial y_\perp^i \partial y_\perp'^j} \right ]_{x_\perp=y_\perp, x_\perp'=y_\perp'}
\!\!\!\!\!\!\! \delta_{\perp}^{ij}
  \\
F_2&=&\int \left [\frac{ \partial^2 C(x_\perp,y_\perp,y_\perp',x_\perp')}{\partial x_\perp^i \partial y_\perp'^j} \right ]_{x_\perp=y_\perp, x_\perp'=y_\perp'}
\!\!\!\!\!\!\! \delta_{\perp}^{ij}
=\int \left [ \frac{ \partial^2 C(x_\perp,y_\perp,y_\perp',x_\perp')}{\partial y_\perp^i \partial y_\perp'^j} \right ]_{x_\perp=y_\perp, x_\perp'=y_\perp'}
\!\!\!\!\!\!\! \delta_{\perp}^{ij}
  \\
F_3&=&
\int \left [ \frac{ \partial^2 C(x_\perp,x_\perp,y_\perp',x_\perp')}{\partial x_\perp^i   \partial x_\perp'^j} \right ]_{ x_\perp'=y_\perp'}
 \delta_{\perp}^{ij}
=\int \left [ \frac{ \partial^2 C(x_\perp,x_\perp,y_\perp',x_\perp')}{\partial y_\perp^i   \partial x_\perp'^j} \right ]_{ x_\perp'=y_\perp'}
 \delta_{\perp}^{ij}
\nonumber \\&=&
\int \left [ \frac{ \partial^2 C(x_\perp,y_\perp,x_\perp',x_\perp')}{\partial y_\perp^i   \partial x_\perp'^j} \right ]_{x_\perp=y_\perp}
 \delta_{\perp}^{ij}
=\int \left [ \frac{ \partial^2 C(x_\perp,y_\perp,x_\perp',x_\perp')}{\partial y_\perp^i   \partial x_\perp'^j} \right ]_{x_\perp=y_\perp}
 \delta_{\perp}^{ij}
\nonumber \\&=& \frac{1}{2}
\int \left [ \frac{ \partial^2  C(x_\perp,x_\perp,x_\perp',x_\perp')}{\partial x_\perp^i   \partial x_\perp'^j} \right ] \delta_{\perp}^{ij} \ ,
\end{eqnarray}
and,
\begin{eqnarray}
H_1&= &\int \left [ \frac{ \partial^2 C(x_\perp,y_\perp,y_\perp',x_\perp')}{\partial x_{\perp,i} \partial x_{\perp,j}' } \right ]_{x_\perp=y_\perp, x_\perp'=y_\perp'}
\bigg( 2\hat k_{1\perp}^i \hat k_{1\perp}^j - \delta_{\perp}^{ij} \bigg)
\nonumber \\&=&
\int \left [ \frac{ \partial^2 C(x_\perp,y_\perp,y_\perp',x_\perp')}{\partial y_\perp^i \partial y_\perp'^j} \right ]_{x_\perp=y_\perp, x_\perp'=y_\perp'}
\bigg( 2\hat k_{1\perp}^i \hat k_{1\perp}^j - \delta_{\perp}^{ij} \bigg)
  \\
H_2&=&\int \left [\frac{ \partial^2 C(x_\perp,y_\perp,y_\perp',x_\perp')}{\partial x_\perp^i \partial y_\perp'^j} \right ]_{x_\perp=y_\perp, x_\perp'=y_\perp'}
\bigg( 2\hat k_{1\perp}^i \hat k_{1\perp}^j - \delta_{\perp}^{ij} \bigg)
\nonumber \\&=&
\int \left [ \frac{ \partial^2 C(x_\perp,y_\perp,y_\perp',x_\perp')}{\partial y_\perp^i \partial y_\perp'^j} \right ]_{x_\perp=y_\perp, x_\perp'=y_\perp'}
\bigg( 2\hat k_{1\perp}^i \hat k_{1\perp}^j - \delta_{\perp}^{ij} \bigg)
  \\
H_3&=&
\int \left [ \frac{ \partial^2 C(x_\perp,x_\perp,y_\perp',x_\perp')}{\partial x_\perp^i   \partial x_\perp'^j} \right ]_{ x_\perp'=y_\perp'}
\bigg( 2\hat k_{1\perp}^i \hat k_{1\perp}^j - \delta_{\perp}^{ij} \bigg)
\nonumber \\&=&
\int \left [ \frac{ \partial^2 C(x_\perp,x_\perp,y_\perp',x_\perp')}{\partial y_\perp^i   \partial x_\perp'^j} \right ]_{ x_\perp'=y_\perp'}
\bigg( 2\hat k_{1\perp}^i \hat k_{1\perp}^j - \delta_{\perp}^{ij} \bigg)
\nonumber \\&=&
\int \left [ \frac{ \partial^2 C(x_\perp,y_\perp,x_\perp',x_\perp')}{\partial y_\perp^i   \partial x_\perp'^j} \right ]_{x_\perp=y_\perp}
\bigg( 2\hat k_{1\perp}^i \hat k_{1\perp}^j - \delta_{\perp}^{ij} \bigg)
\nonumber \\&=&
\int \left [ \frac{ \partial^2 C(x_\perp,y_\perp,x_\perp',x_\perp')}{\partial y_\perp^i   \partial x_\perp'^j} \right ]_{x_\perp=y_\perp}
\bigg( 2\hat k_{1\perp}^i \hat k_{1\perp}^j - \delta_{\perp}^{ij} \bigg)
\nonumber \\&=& \frac{1}{2}
\int \left [ \frac{ \partial^2  C(x_\perp,x_\perp,x_\perp',x_\perp')}{\partial x_\perp^i   \partial x_\perp'^j} \right ]
\bigg( 2\hat k_{1\perp}^i \hat k_{1\perp}^j - \delta_{\perp}^{ij} \bigg) \ ,
\end{eqnarray}
with,
\begin{eqnarray}
F_1&=&2 \pi^4 N_c \alpha_s x_1\left [ G_{DP}(x_1,k_{1\perp})+\left (1-\frac{4}{N_c^2} \right )  G_{q\bar q}(x_1,k_{1\perp})
+\frac{2}{N_c^2} G_{WW}(x_1,k_{1\perp}) \right ]
\\
F_2&=&2 \pi^4 N_c \alpha_s x_1\left [ G_{DP}(x_1,k_{1\perp})-\left (1-\frac{4}{N_c^2} \right )  G_{q\bar q}(x_1,k_{1\perp})
-\frac{2}{N_c^2} G_{WW}(x_1,k_{1\perp}) \right ]
\\
F_3&=&2 \pi^4 N_c \alpha_s x_1\left [ 2G_{DP}(x_1,k_{1\perp}) \right ]
\\
H_1&=&2 \pi^4 N_c \alpha_s x_1 \left [ h_{1,DP}^{\perp g}(x_1,k_{1\perp})+\left (1-\frac{4}{N_c^2} \right )  h_{1,q\bar q}^{\perp g}(x_1,k_{1\perp})
+\frac{2}{N_c^2} h_{1,WW}^{\perp g}(x_1,k_{1\perp}) \right ]
\\
H_2&=&2 \pi^4 N_c \alpha_s x_1 \left [ h_{1,DP}^{\perp g}(x_1,k_{1\perp})-\left (1-\frac{4}{N_c^2} \right )  h_{1,q\bar q}^{\perp g}(x_1,k_{1\perp})
-\frac{2}{N_c^2} h_{1,WW}^{\perp g}(x_1,k_{1\perp}) \right ]
\\
H_3&=&2 \pi^4 N_c \alpha_s x_1 \left [ 2 h_{1,DP}^{\perp g}(x_1,k_{1\perp}) \right ] \ .
\end{eqnarray}
To arrive at the results given above, we have neglected the logarithmic dependence of the saturation momentum on $r_\perp^2$.
$G_{DP}, G_{WW}, h_{1,DP}^{\perp g}$ and $h_{1,WW}^{\perp g}$ are the  unpolarized  gluon dipole distribution, the  Weizs\"{a}cker-Williams (WW) type
 unpolarized gluon distribution, the dipole type linearly polarized gluon distribution,
 and the WW type linearly polarized gluon distribution, respectively.
 In the MV model, they read~\cite{Kovchegov:1996ty,JalilianMarian:1996xn,Metz:2011wb},
\begin{eqnarray}
 x_1 G_{DP}(x_1, k_{1\perp})\!\!\! & = & \!\!\!x_1 h^{\perp g}_{1,DP}(x_1,k_{1\perp})=
\frac{C_F S_\perp}{ 2 \pi^2 \alpha_s}k_{1\perp}^2
\int \frac{d^2 r_\perp}{(2\pi)^2} \, e^{-i  k_{1\perp} \cdot  r_\perp}
e^{-\frac{r_\perp^2 Q_{s}^2 }{4}}
\\
x_1 G_{WW}(x_1,k_{1\perp}) \!\!\!
& = & \!\!\! \frac{N_c^2-1 }{N_c} \frac{S_\perp }{4 \pi^4 \alpha_s} \int d^2 r_\perp
e^{-i  k_{1\perp} \cdot  r_\perp} \,
\frac{1}{r_\perp^2} \bigg( 1 - e^{ - \frac{r_\perp^2 Q_s^2}{4}} \bigg)
\\
x_1 h^{\perp g}_{1,WW}(x_1,k_{1\perp}) \!\!\! &=& \!\!\!
\frac{N_c^2-1 }{8 \pi^3} \, S_\perp \int d r_\perp \,
\frac{J_2 (|k_{1\perp}|| r_\perp)|}{\frac{1}{4 \mu_A} |r_{\perp}| Q_s^2}
\bigg( 1 - e^{ - \frac{r_\perp^2 Q_s^2}{4}} \bigg) \,.
\label{eq:1}
\end{eqnarray}
Here $S_\perp$ is the transverse area of the target nucleus.
$Q_s^2 = \alpha_s N_c \mu_A{\rm ln} \left [ 1 / (r_\perp^2 \Lambda_{QCD}^2) \right ]$ is the
gluon saturation scale with $\mu_A$ being a common CGC parameter. $J_2$ is the second order Bessel function.
Note that our convention for $h^{\perp g}_{1,WW}$  differs from that in
Ref.~\cite{Metz:2011wb} by a factor 1/2.
The WW type gluon distributions have a clear physical interpretation as the number density of gluons inside a hadron/nucleus,
while the dipole type distribution does not.  On the other hand, the
dipole type unpolarized gluon distribution in the adjoint representation
 enters the single gluon production cross section  in pA collisions~\cite{Kovchegov:1998bi}.
Besides these widely used gluon TMDs, two novel ones are given by,
 \begin{eqnarray}
 x_1 G_{q \bar q}(x_1, k_{1\perp})\!\!\! & = & \!\!\!
\frac{C_F S_\perp}{ 2 \pi^2 \alpha_s}
\int \frac{d^2 r_\perp}{(2\pi)^2}  \, e^{-i  k_{1\perp} \cdot  r_\perp} \,
Q_s^2 e^{-\frac{r_\perp^2 Q_{s}^2 }{4}} \ ,
\\
x_1 h^{\perp g}_{1,q \bar q}(x_1,k_{1\perp})\!\!\! & = & \!\!\!
\frac{N_c^2-1 }{8 \pi^3} \, S_\perp \int d |r_\perp| \mu_A |r_\perp|
J_2 (|k_{1\perp}|| r_\perp|)
 e^{ - \frac{r_\perp^2 Q_s^2}{4}}  \, .
\end{eqnarray}

Collecting all the  pieces together, the differential cross section for quark pair production can be written in the following general form,
\begin{eqnarray}
\frac{d \sigma}{d{\cal P.S.}}=\frac{\alpha_s^2 N_c}{\hat s^2 (N_c^2-1)}
\left [ {\cal A}(q_\perp^2)+  \frac{m^2}{P_\perp^2} {\cal B}(q_\perp^2) \cos 2\phi +{\cal C}(q_\perp^2)\cos 4\phi \right ]
\end{eqnarray}
where $\phi$ is the azimuthal angle between the transverse momenta $ q_\perp$ and $ P_\perp$. The coefficients
${\cal A}(q_\perp^2)$, $ {\cal B}(q_\perp^2)$ and ${\cal C}(q_\perp^2)$ contain convolutions of various gluon TMDs.
Instead of presenting the full results for these coefficients, we neglect all  higher powers in $m^2/P_\perp^2$ ,
\begin{eqnarray}
&& \!\!\!\!\!\!\!\!\!\!\!\!\!\!
{\cal A}(q_\perp^2)=\int  d^2 k_{1\perp}d^2 k_{2\perp}  \delta^2(k_{1\perp}+k_{2\perp}-q_\perp) x_2 g (x_2,k_{2\perp} )
  \frac{(\hat u^2+\hat t^2)}{4\hat u \hat t}
\nonumber \\ && \times \left \{ \frac{(\hat t- \hat u)^2}{\hat s^2}  x_1 G_{DP}(x_1, k_{1\perp})+
 x_1 \left [ \left (1-\frac{4}{N_c^2} \right )G_{q\bar q}(x_1,k_{1\perp})
+\frac{2}{N_c^2} G_{WW}(x_1,k_{1\perp}) \right ] \right \}
 \\
 \nonumber \\
&&  \!\!\!\!\!\!\!\!\!\!\!\!\!\!
{\cal B}(q_\perp^2)=\int  d^2 k_{1\perp}d^2 k_{2\perp}  \delta^2(k_{1\perp}+k_{2\perp}-q_\perp) x_2 g (x_2,k_{2\perp} )
\nonumber \\ & & \times
 \left \{ \left [2(\hat k_{1\perp} \cdot \hat q_\perp)^2-1 \right ] \right .\
\nonumber \\ &  & \left .\ \times
 \left ( \frac{(\hat t -\hat u)^2}{\hat s^2} x_1h_{1,DP}^{\perp g}(x_1,k_{1\perp})
 + x_1 \left [ \left (1-\frac{4}{N_c^2} \right )  h_{1,q\bar q}^{\perp g}(x_1,k_{1\perp})
+\frac{2}{N_c^2} h_{1,WW}^{\perp g}(x_1,k_{1\perp})  \right ] \right )  \right .\
 \nonumber \\ && \ \ \ \  +
 \left [2(\hat k_{2\perp} \cdot \hat q_\perp)^2-1 \right ]
  \nonumber \\ && \left .\ \times
  \left ( \frac{(\hat t -\hat u)^2}{\hat s^2} x_1G_{DP}(x_1,k_{1\perp})
 +x_1 \left [ \left (1-\frac{4}{N_c^2} \right )G_{q\bar q}(x_1,k_{1\perp})
+\frac{2}{N_c^2} G_{WW}(x_1,k_{1\perp}) \right ] \right ) \right \}
 \\
 \nonumber \\
&&  \!\!\!\!\!\!\!\!\!\!\!\!\!\!
{\cal C}(q_\perp^2)=\int  d^2 k_{1\perp}d^2 k_{2\perp}  \delta^2(k_{1\perp}+k_{2\perp}-q_\perp) x_2 g (x_2,k_{2\perp} )
\nonumber \\ && \times
\left [   \left (2(\hat q \cdot \hat k_{1\perp})(\hat q \cdot \hat  k_{2\perp})-\hat k_{1\perp} \cdot \hat k_{2\perp} \right )^2-\frac{1}{2} \right ]
\nonumber \\
 & & \times \left \{   \frac{(\hat t -\hat u)^2}{\hat s^2} x_1h_{1,DP}^{\perp g}(x_1,k_{1\perp})  +
x_1\left [ \left (1-\frac{4}{N_c^2} \right )  h_{1,q\bar q}^{\perp g}(x_1,k_{1\perp})
+\frac{2}{N_c^2} h_{1,WW}^{\perp g}(x_1,k_{1\perp}) \right ] \right \} \ ,
\end{eqnarray}
where $\hat s=(x_1 P_A +x_2 P_B)^2$, $\hat t=(x_2 P_B -l_1)^2$ and $\hat u=(x_1 P_A-l_1)^2$ are kinematical variables
defined in the usual way. This is the main result of our paper.

A few remarks are in order on the above analytical result.
\begin{itemize}
\item
One notices that six different  types of TMD gluon distributions are
involved in the azimuthal angle dependent differential cross section,
among which three are unpolarized gluon TMDs and the rest are linearly polarized gluon distributions.
They differ due to the different gauge link structures arising from initial/final state interaction(ISF/FSF).
Thus, by  measuring the di-jet imbalance and the azimuthal asymmetries one can investigate
how the gluon transverse momentum spectrum is affected by ISF/FSF.
\item
We have taken into account the $N_c$ suppressed terms in both the
unpolarized and polarized cross sections.  As discussed in the next section,
the $N_c$ suppressed terms play an important role for low transverse momentum.
Therefore, the large $N_c$ limit adopted in the papers~\cite{Dominguez:2010xd,Dominguez:2011wm}
is actually not a good approximation in certain kinematical regions.
\item
The $\cos 2 \phi$ azimuthal asymmetry is proportional to the mass of the produced quark.
Therefore, it might be optimal to study this asymmetry for charm and bottom quark-antiquark
pair production at RHIC and LHC.
\item
It is worthwhile to point out that as observed in~\cite{Sun:2011iw}, one automatically
takes into account the contribution from the linearly
polarized gluon TMD in $k_t$ factorization. In other words,
the usual unpolarized gluon distribution of the nucleon is the same as its linearly polarized
gluon distribution in the Lipatov approximation.
\item
Finally, we would like to mention that
it is also feasible to take into account small $x$ evolution effect~\cite{JalilianMarian:1997gr,Dumitru:2010mv}.
\end{itemize}

\section{The dilute limit, forward limit, and large $N_c$ limit}
In this section, we show how the obtained complete  analytical results reduce to the existing  results in the literatures in the dilute limit, and large $N_c$ limit in the
nucleon forward region.

We first discuss the expression in the dilute limit.
In the correlation limit, the low gluon densities limit is reached in the kinematic region $Q_s^2 \ll k_{1\perp}^2 \ll P_\perp^2$.
 When $Q_s^2 \ll k_{1\perp}^2$ , all six gluon distribution functions become identical,
though they differ significantly at low $k_{1\perp}$,
\begin{eqnarray}
&& \!\!\!\!\!\!\!\!\!\!\!\!\!\!\!\!\!\!\!\!\!\!\!\!\!\!\!\!\!\!\!\!\!\!\!
x_1G(x_1,k_{1\perp})  \equiv
 x_1h_{1,DP}^{\perp g}(x_1,k_{1\perp}) =x_1 h_{1,WW}^{\perp g}(x_1,k_{1\perp})= x_1h_{1,q\bar q}^{\perp g}(x_1,k_{1\perp})
 \nonumber \\
&& = x_1G_{DP}(x_1,k_{1\perp})=x_1G_{WW}(x_1,k_{1\perp})= x_1G_{q\bar q}(x_1,k_{1\perp})
 \simeq S_\perp \frac{N_c^2-1}{4 \pi^3}\frac{\mu_A}{k_{1\perp}^2}
 \, .
\end{eqnarray}
Note that the well known bremsstrahlung spectrum $1/k_{1\perp}^2$ is recovered for all types of gluon TMD distributions in the dilute limit.
This is because, when the gluon densities of the nuclear target are not too large, the multiple gluon re-scattering plays a less important role in describing
the gluon transverse momentum spectrum.
By replacing the various  gluon distributions appearing in the coefficients ${\cal A}(q_\perp^2)$, $ {\cal B}(q_\perp^2)$ and  $  {\cal C}(q_\perp^2)$
 with the above dilute gluon distribution, we have,
\begin{eqnarray}
{\cal A}(q_\perp^2) &=& \int  d^2 k_{1\perp}d^2 k_{2\perp}  \delta^2(k_{1\perp}+k_{2\perp}-q_\perp) x_2 g (x_2,k_{2\perp} )x_1G(x_1,k_{1\perp})
 \nonumber \\ &  & \times
  \left [ \frac{N_c^2-1}{2N_c^2} \frac{\hat u^2+\hat t^2}{\hat u \hat t}  - \frac{\hat t^2 + \hat u^2}{\hat s^2} \right ]
 \\
 \nonumber \\
{\cal B}(q_\perp^2)&=& \int  d^2 k_{1\perp}d^2 k_{2\perp}  \delta^2(k_{1\perp}+k_{2\perp}-q_\perp) x_2 g (x_2,k_{2\perp} )x_1G(x_1,k_{1\perp})
\nonumber \\ &  & \times
 4 \left [ \frac{N_c^2-1}{2N_c^2}-\frac{\hat t \hat u}{\hat s^2} \right ]
  \left [ 2(\hat k_{1\perp} \cdot \hat q_\perp)^2 +  2(\hat k_{2\perp} \cdot \hat q_\perp)^2-2   \right ]
 \\
 \nonumber \\
{\cal C}(q_\perp^2)&=&\int  d^2 k_{1\perp}d^2 k_{2\perp}  \delta^2(k_{1\perp}+k_{2\perp}-q_\perp) x_2 g (x_2,k_{2\perp} )x_1G(x_1,k_{1\perp})
\nonumber \\ && \times
 4 \left [ \frac{N_c^2-1}{2N_c^2}-\frac{\hat t \hat u}{\hat s^2} \right ]
\left [   \left (2(\hat q \cdot \hat k_{1\perp})(\hat q \cdot \hat  k_{2\perp})-\hat k_{1\perp} \cdot \hat k_{2\perp} \right )^2-\frac{1}{2} \right ] \ .
\end{eqnarray}
Here the known unpolarized Born cross section for $q \bar q$ production through gluon fusion has been recovered
for the unpolarized term, as it should be.
Agreement is also found between our results and the explicit expressions of the polarized cross section given
in~\cite{Boer:2009nc,Boer:2010zf},
provided that these results are extended to the small $x$ region and the same dilute limit is taken.
As mentioned in the previous section, one automatically takes into account the linearly polarized
gluons inside a proton in the Lipatov approximation.
The result presented in~\cite{Boer:2009nc,Boer:2010zf} were computed in the TMD factorization approach.
In principle, the gluon TMDs associated with different hard scattering processes contain different
gauge link structures.
However, the non-trivial initial/final state interaction effects encoded in the gauge links were not
quantitatively analyzed in~\cite{Boer:2009nc,Boer:2010zf}.
Therefore, by observing these consistences, we conclude that
in the dilute limit, the contribution from initial/final state interactions encoded in the gauge links
can be neglected, and single gluon exchange dominates the processes.

Let us now discuss the expressions we obtain in the nucleon forward limit.
Since the gluon intrinsic transverse momentum $k_{2\perp}$ inside a nucleon
can be neglected in the forward limit as compared to that
from the gluon distribution of nucleus, we may make
the approximation $ \delta^2(k_{1\perp}+k_{2\perp}-q_\perp) \approx  \delta^2(k_{1\perp}-q_\perp)$
and integrate out $k_{1\perp}$ and $k_{2\perp}$.
In doing so, we essentially recover a hybrid approach widely used in the CGC calculation,
in which one applies the collinear factorization for the
integrated gluon or quark distributions inside the dilute proton at large $x_2$,
while the CGC formalism is used on the nucleus side.
After making such approximations, one ends up with the following simplified result,
\begin{eqnarray}
&& \!\!\!\!\!\!\!\!\!\!\!\!\!\!
{\cal A}(q_\perp^2)= x_2 g (x_2)\frac{(\hat u^2+\hat t^2)}{4 \hat u \hat t}
 \nonumber \\&&
\times  \left \{ \frac{(\hat t- \hat u)^2}{\hat s^2}  x_1 G_{DP}(x_1, q_\perp)+
 x_1 \left [ \left (1-\frac{4}{N_c^2} \right ) G_{q \bar q}(x_1,q_\perp)
+\frac{2}{N_c^2} G_{WW}(x_1,q_\perp) \right ] \right \}
\\
 \nonumber \\
&&  \!\!\!\!\!\!\!\!\!\!\!\!\!\!
{\cal B}(q_\perp^2)= x_2 g (x_2 )
 \nonumber \\&& \times
  \left \{ \frac{(\hat t -\hat u)^2}{\hat s^2} x_1h_{1,DP}^{\perp g}(x_1,q_\perp)
+x_1 \left [ \left (1-\frac{4}{N_c^2} \right ) h_{1,q \bar q}^{\perp g}(x_1,q_\perp)
 +\frac{2}{N_c^2}  h_{1,WW}^{\perp g}(x_1,q_\perp) \right ] \right \}
\\
\nonumber \\
&&  \!\!\!\!\!\!\!\!\!\!\!\!\!\!
{\cal C}(q_\perp^2)=0  \ ,
\end{eqnarray}
where $g(x_2)$ is the integrated gluon distribution function for the proton. It is shown that
the $\cos 4 \phi$ modulation arising from the
product of two linearly polarized gluon distributions from both the nucleon and
nucleus drops out in the forward limit.
This is because the linearly polarized gluon distribution of
the nucleon disappears after integrating over the gluon transverse momentum.

In order to compare with existing results for quark pair production in $pA$ collisions
in the forward limit, we further take the large $N_c$ limit,
\begin{eqnarray}
&& \!\!\!\!\!\!\!\!\!\!\!\!\!\!
{\cal A}(q_\perp^2)= x_2 g (x_2)
\frac{(\hat u^2+\hat t^2)}{4\hat u \hat t}
\left \{ \frac{(\hat t- \hat u)^2}{\hat s^2}  x_1 G_{DP}(x_1, q_\perp)+
 x_1  G_{q \bar q}(x_1,q_\perp)  \right \}
 \\
&&  \!\!\!\!\!\!\!\!\!\!\!\!\!\!
{\cal B}(q_\perp^2)= x_2 g (x_2 )
  \left \{ \frac{(\hat t -\hat u)^2}{\hat s^2} x_1h_{1,DP}^{\perp g}(x_1,q_\perp)
+x_1  h_{1,q \bar q}^{\perp g}(x_1,q_\perp) \right \}
 \\
&&  \!\!\!\!\!\!\!\!\!\!\!\!\!\!
{\cal C}(q_\perp^2)=0 \ ,
\end{eqnarray}
where the unpolarized cross section is in agreement with that obtained in Ref.~\cite{Dominguez:2011wm}
if one uses the relations
$ x_1 G_{DP}(x_1, q_\perp)={\cal F}_{gg}^{(1)}-{\cal F}_{gg}^{(2)}$ and $  x_1  G_{q \bar q}(x_1,q_\perp)={\cal F}_{gg}^{(1)}+{\cal F}_{gg}^{(2)}$
valid in the large $N_c$ limit. ${\cal F}_{gg}^{(1)}$ and ${\cal F}_{gg}^{(2)}$ are expressed as
convolution between the dipole gluon distribution and a
Gaussian form and are given by~\cite{Dominguez:2011wm},
\begin{eqnarray}
{\cal F}_{gg}^{(1)}(x_1, q_\perp)&=& \int d^2 q_{1\perp} d^2 q_{2\perp} \delta^2( q_\perp-q_{1\perp}- q_{2\perp})
x_1 G_{DP}(x_1, q_{1\perp}) F(q_2) \ ,
 \\
{\cal F}_{gg}^{(2)}(x_1, q_\perp) &=&- \int d^2 q_{1\perp} d^2 q_{2\perp} \delta^2( q_\perp-q_{1\perp}- q_{2\perp})
\frac{q_{1\perp} \cdot  q_{2\perp}}{q_{1\perp} ^2}
x_1 G_{DP}(x_1, q_{1\perp}) F(q_2)
  \ ,
\end{eqnarray}
where $F(q_2)$ is a Gaussian and its definition can be found in Ref.~\cite{Dominguez:2011wm}.
At this point, we would like to emphasize that the large $N_c$
limit is not necessarily a good approximation.
In particular, $N_c$ suppressed terms could be the dominant
contribution in the dense medium region.
This can be best seen by investigating how the various gluon TMDs involved
in unpolarized and polarized cross sections scale as at low $k_{1\perp}$:
 \begin{eqnarray}
&& x_1G_{DP}(x_1, k_{1\perp})=x_1 h_{1,DP}^{\perp g}(x_1,k_{1\perp})\sim k_{1\perp}^2/Q_s^2,
 \ \ \ \ x_1G_{q \bar q}(x_1,k_{1\perp}) \sim constant ,
 \nonumber \\
&& x_1G_{WW}(x_1,k_{1\perp}) \sim {\rm ln}(Q_s^2/k_{1\perp}^2), \ \ \ \
x_1 h_{1,WW}^{\perp g}(x_1,k_{1\perp}) \sim x_1 h_{1,q \bar q}^{\perp g}(x_1,k_{1\perp}) \sim \mu_A/Q_s^2
  \ .
\end{eqnarray}
Clearly, the term proportional to the WW type unpolarized gluon distribution is the dominant
one in the unpolarized differential cross section at low transverse momentum,
as it  keeps rising like the logarithm of $1/k_{1\perp}^2$ in the saturation regime
where all other gluon distributions either approach a constant or vanish.
In contrast to the unpolarized case, the sub-leading $N_c$ contribution is indeed suppressed by a factor $2/N_c^2$
as compared to the leading $N_c$ contribution in the $\cos 2\phi$ dependent differential
cross section at low transverse momentum.

\section{Quark pair production in TMD factorization}
Roughly speaking, transverse momentum dependent factorization
applies in the hard scattering processes when a hard scale involved
in the corresponding processes is much larger than the
parton intrinsic transverse momenta. This is indeed the case for
quark pair production in pA collisions in the correlation limit
where the individual quark transverse momentum serves as the hard
scale of the process and is much larger than the transverse momentum
imbalance of the quark pair related to incoming gluon transverse
momenta. In general, the differential cross section computed in the
TMD factorization framework can be factorized into the hard partonic
cross section and the various spin and transverse momentum dependent
parton distributions. Hard parts are perturbatively calculable,
while the parton TMDs are normally regarded as universal
non-perturbative objects. The proper gauge invariant definitions of
TMDs involve nonlocal operators containing path-ordered
exponentials, the gauge links, which result from resumming  all
longitudinally polarized gluons into the soft parts. The gauge link
has important physical effects, and particularly plays a central
role in the description of transverse single spin asymmetries as
well as transverse momentum broadening in high energy collisions
involving  a large nucleus~\cite{Liang:2008vz}.

However, it has been realized that standard TMD factorization fails in
di-jet production in hadronic collisions~\cite{Collins:2007nk}.
Since the structure of gauge links generally depends on the process,
TMD distributions are essentially process dependent,
implying a breakdown of universality. A solution to this problem has been proposed by
introducing the so-called generalized TMD factorization~\cite{Bomhof:2004aw},
in which the basic factorized structure is assumed to remain valid, but with TMD
distributions that contain non-standard, process dependent gauge link
structures. In the framework of generalized TMD factorization, the modified gauge links
are obtained by resumming longitudinally polarized gluons into
parton correlation functions on each nucleon side separately.
However, recent work has shown that it is impossible to do so for di-jet production
in nucleon nucleon collisions
because the initial/final state interaction will not allow a separation of
gauge links into the matrix elements of the various TMDs associated with each
incoming hadron.
This has been explicitly illustrated by a concrete counter-example in Ref.~\cite{Rogers:2010dm}.

Similarly, for quark pair production in hadronic collisions, generalized TMD factorization is not valid any longer.
However, in pA collisions, if one only takes into account the interaction between the active partons and
the background gluon field inside a large nucleus while
neglecting the longitudinal gluons attached to the proton side, the type of graph
(for example Fig.11 in~\cite{Rogers:2010dm})
which can produce a violation of generalized TMD-factorization disappears.
In this section, we use this approximation. Admittedly we cannot quantify the systematic errors introduced by it.
After neglecting the extra gluon attachment on the proton side, the multiple gluon re-scattering between
the hard part and the nucleus can be resummed to all orders in the form
of a process dependent gauge link.
Due to the different color structures, the gluon TMDs associated with different Feynman diagrams
correspond to different gauge link structures.
For example, the gluon TMD correlation function associated with graph Fig.3(a) takes
the form~\cite{Bomhof:2004aw},
\begin{eqnarray}
\Phi_{g,(a)}^{ij}&=&2 \int \frac{d \xi^- d \xi_\perp}{(2\pi)^3 P^+_A} e^{ix_1P_A^+-ik_{1\perp} \cdot \xi_\perp}
\nonumber \\ & & \times \langle P|
{\rm  Tr_c} \left \{    F^i(\xi)  \left [  \frac{N_c^2}{N_c^2-1}   \frac{ {\rm Tr} \left [ U^{[ \!\!\!\!\!\!\!\! \qed ]\dag} \right ] }{N_c}
U^{[-]\dag}- \frac{1}{N_c^2-1} U^{[+]\dag} \right ] F^j (0) U^{[+]}     \right \} |P \rangle \ ,
\end{eqnarray}
where $i,j$ denote the gluon polarization index. The gauge links $U^{[+]}$, $U^{[-]}$ are defined as,
\begin{eqnarray}
U^{[+]}&=& \mathcal{P} \, e^{-ig_s \int_{0}^{\infty} d \zeta^- A^+(\zeta^-, 0_{\perp})} \,
\mathcal{P} \, e^{-ig_s \int^{\xi^-}_{\infty}  d \zeta^- A^+(\zeta^-, \xi_{\perp})} \,,
\\
U^{[-]}&=& \mathcal{P} \, e^{-ig_s \int_{0}^{-\infty} d \zeta^- A^+(\zeta^-, 0_{\perp})} \,
\mathcal{P} \, e^{-ig_s \int^{\xi^-}_{-\infty}  d \zeta^- A^+(\zeta^-, \xi_{\perp})} \, .
\end{eqnarray}
 And $U^{[ \!\!\!\!\!\!\!\! \qed ]}=U^{[+]}U^{[-]\dag}=U^{[-]\dag} U^{[+]}$ emerges as a Wilson loop.
 At small $x$, this gluon TMD can be expressed as the derivative of a
multiple-point function and subsequently be computed in the MV model.
In order to derive this relation, we make use of the Fierz identities,
\begin{eqnarray}
&& \!\!\!\!\!\!\!\!\!\!
C(x_\perp,y_\perp,y_\perp',x_\perp')={\rm Tr_c} \langle U(x_\perp) t^a U^\dag(y_\perp) U(y_\perp') t^a U^\dag(x_\perp')   \rangle
 \nonumber \\
&&=\frac{1}{2} {\rm Tr_c} \langle  U^\dag(x_\perp') U(x_\perp) \rangle {\rm Tr_c} \langle U^\dag(y_\perp) U(y_\perp') \rangle
-\frac{1}{2N_c}{\rm Tr_c} \langle  U(x_\perp)  U^\dag(y_\perp) U(y_\perp')  U^\dag(x_\perp') \rangle ,
\end{eqnarray}
and the formula,
\begin{eqnarray}
\partial^i U(x_\perp)=-i g_s \int_{-\infty}^\infty dx^- U[-\infty,x^-,x_\perp] \partial^iA^+(x^-,x_\perp) U[x^-,\infty,x_\perp] \ .
\end{eqnarray}
With the help of the above two identities, one finds,
\begin{eqnarray}
\Phi_{(a)}^{ij}=  \frac{2N_c}{N_c^2-1} \frac{2}{\alpha_s}\int \frac{d^2x_\perp d^2 x_\perp'}{(2\pi)^4} e^{-ik_{1\perp}(x_\perp-x_\perp')}
\left [ \frac{\partial^2}{\partial x_{\perp,i} \partial x_{\perp,j}'}C(x_\perp,y_\perp,y_\perp',x_\perp') \right ]_{x_\perp=y_\perp, \  x_\perp'=y_\perp'}
\end{eqnarray}
The normalization on the right hand side of the
equation is fixed according to the arguments made in Ref.\cite{Dominguez:2011wm}.
Following a similar procedure, for the gluon distribution correlation functions associated with other diagrams shown in Fig.3,
we obtain,
\begin{eqnarray}
\Phi_{(b)}^{ij}&=& \frac{2N_c}{N_c^2-1} \frac{2}{\alpha_s}\int \frac{d^2x_\perp d^2 x_\perp'}{(2\pi)^4} e^{-ik_{1\perp}(x_\perp-x_\perp')}
\left [ \frac{\partial^2}{\partial y_{\perp,i} \partial y_{\perp,j}'}C(x_\perp,y_\perp,y_\perp',x_\perp') \right ]_{x_\perp=y_\perp, \  x_\perp'=y_\perp'}
\\
\Phi_{(c)}^{ij}&=&  2N_c \frac{2}{\alpha_s}\int \frac{d^2x_\perp d^2 x_\perp'}{(2\pi)^4} e^{-ik_{1\perp}(x_\perp-x_\perp')}
\left [ \frac{\partial^2}{\partial x_{\perp,i} \partial y_{\perp,j}'}C(x_\perp,y_\perp,y_\perp',x_\perp') \right ]_{x_\perp=y_\perp, \  x_\perp'=y_\perp'}
\\
\Phi_{(d)}^{ij}&=& \frac{1}{N_c} \frac{2}{\alpha_s}\int \frac{d^2x_\perp d^2 x_\perp'}{(2\pi)^4} e^{-ik_{1\perp}(x_\perp-x_\perp')}
\left [ \frac{\partial^2}{\partial x_{\perp,i} \partial x_{\perp,j}'}C(x_\perp,x_\perp,x_\perp',x_\perp') \right ]
\\
\Phi_{(e)}^{ij}&=&  \frac{2}{N_c} \frac{2}{\alpha_s}\int \frac{d^2x_\perp d^2 x_\perp'}{(2\pi)^4} e^{-ik_{1\perp}(x_\perp-x_\perp')}
\left [ \frac{\partial^2}{\partial x_{\perp,i} \partial x_{\perp,j}'}C(x_\perp,x_\perp,y_\perp',x_\perp') \right ]_{ \  x_\perp'=y_\perp'}
\\
\Phi_{(f)}^{ij}&=& \frac{2}{N_c}  \frac{2}{\alpha_s}\int \frac{d^2x_\perp d^2 x_\perp'}{(2\pi)^4} e^{-ik_{1\perp}(x_\perp-x_\perp')}
\left [ \frac{\partial^2}{\partial x_{\perp,i} \partial y_{\perp,j}'}C(x_\perp,x_\perp,y_\perp',x_\perp') \right ]_{  x_\perp'=y_\perp'}
\end{eqnarray}
With these relations,  all of the unpolarized and linearly polarized gluon TMDs can be calculated in the MV model.
On the other hand, it is straightforward to compute the partonic hard cross section contributions from each diagram in the Fig.3.
Combining the derived gluon TMDs and hard parts and summing the contributions from all diagrams,
we obtain the finial result in the TMD factorization framework.
In order to compare the obtained TMD factorization result with that calculated in the CGC formalism,
we have to take the same dilute limit on the proton side, which means
that the unpolarized gluon distribution and the linearly polarized gluon distribution
inside a proton become identical.  After making this assumption, a perfect matching
between the CGC formalism and TMD factorization is found in the correlation limit. We emphasize  that this conclusion is valid beyond the large $N_c$ limit.
\begin{figure}[t]
\begin{center}
\includegraphics[width=13cm]{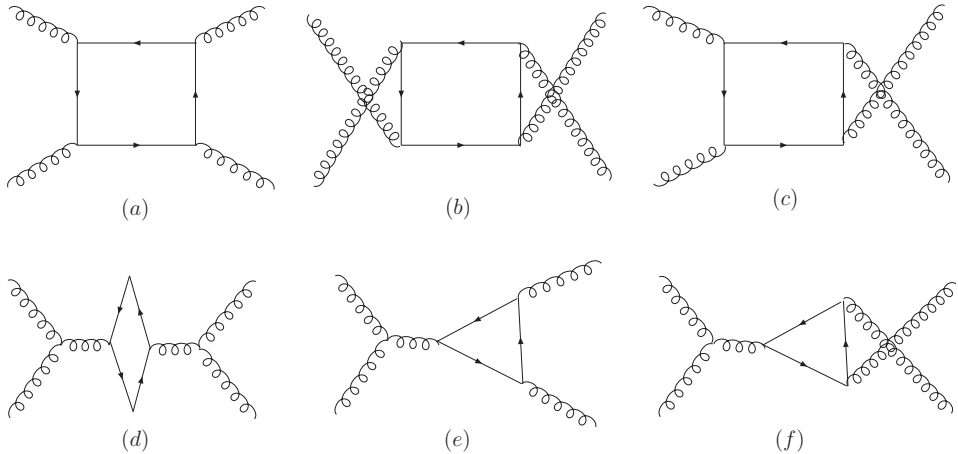}
\end{center}
\vskip -0.4cm \caption{ The diagrams contributing to quark pair production in TMD factorization approach.
The gauge link structure of gluon TMD distributions associated with each diagram are different.
The mirror diagrams are not shown.}
\label{efig2}
\end{figure}

As an effective TMD factorization is established in the quark pair
production process, the measurement of the azimuthal asymmetries in
pA collisions allows to probe  directly the distribution of the
linearly polarized gluons inside a large nucleus.  Such measurements
provide us with a first chance to explore the gluon polarization
effect in the saturation regime. Since the magnitude of various
linearly polarized gluon distributions are of the same magnitude as
the unpolarized ones at small $x$ ( this becomes evident in the
dilute limit where polarized and unpolarized distributions become
identical), we also anticipate that these asymmetries are quite
sizeable, suggesting  a promising prospect for the extraction of
polarized gluon distributions from quark pair production process. By
comparing the gluon distributions extracted from this process with
that probed in the other processes, like di-jet production in  eA
collisions, one could deduce how the gluon transverse momentum
spectrum is affected by the different initial/final state
interactions.

To emphasize the phenomenological relevance of our results, let us add that recently
a strong back-to-back de-correlation of the two hadrons in $dAu$
collisions in the forward rapidity region of the deuteron was
discovered by STAR and PHENIX~\cite{Braidot:2010ig,Adare:2011sc}.
However, at RHIC energy, the dominant channel is $qg\rightarrow qg$  in the forward region.
The $gg\rightarrow q\bar q$ channel only becomes relevant in the central rapidity region at RHIC.
Apart from this, the effects caused by the polarized gluon distributions
can not be isolated by only looking at the angular deviation from the back-to-back situation,
but depend on the jet transverse energy~\cite{Boer:2009nc}.
Finally, it is important to mention that the small $x$ evolution effect has to be taken
into account at LHC since the MV model is only a good model for
high energy scattering when $x$ is not smaller than $0.01$ for a large nucleus.
It should be feasible  to measure these polarization dependent observables at RHIC and LHC.
We plan to perform a complete set of phenomenological studies to investigate this
possibility in the future.

\section{Summary}
\noindent

In this paper, we have studied quark pair production in high energy
proton-nucleus collisions in the central rapidity region and in the
correlation limit where the total transverse momentum of the quark
pair ($q_\perp$) is much smaller than the transverse momenta of the
individual quarks ($\approx P_\perp$). Our main focus lay on the
polarized case. We first used a hybrid approach to reproduce the
full CGC result for quark pair production beyond the correlation
limit. Our hybrid approach allowed us to take into account finite
gluon transverse momentum effects on the proton side in a certain
approximation. Employing a power expansion in the correlation limit,
the multiple-point functions appearing in the full CGC result
collapse into two-point functions and are thus given by a
combination of gluon TMDs. All finite $N_c$ terms are kept in our
calculation. The resulting cross section contains $\cos 2\phi$ and
$\cos 4 \phi$ dependent terms, where $\phi$ is the azimuthal angle
between the transverse momenta $q_\perp$ and $ P_\perp$. In addition
to WW and dipole type linearly polarized gluon distributions, the
novel linearly polarized gluon distribution $ h^{\perp g}_{1,q \bar
q}$ also generates $\cos 2\phi$ and $\cos 4 \phi$ modulations. Such
asymmetries could be measured at RHIC and LHC.

We further discussed our results in the dilute limit, the forward limit and the large $N_c$ limit,
and found consistency with existing results in the different limits. The physical implications of
the observed consistences were also addressed.
In the end, we showed that a calculation based on TMD factorization leads to the same result as that
obtained in the hybrid approach.
The technique introduced in this paper can be extended to study di-jet production in other channels
for pA collisions (e.g. di-jets initiated by different partons and/or various polarization channels).
For these the linearly polarized gluon TMDs with different gauge link structures
may also manifest themselves through $\cos 2\phi$ or $\cos 4\phi$ azimuthal dependencies of the cross sections.
We would expect that exploring these polarization obervables at small $x$ will open a new path
to investigate spin physics as well as saturation physics.\\

{\bf Acknowledgments:}
One of us (Jian Zhou) thanks Andreas Metz for suggesting this work to him and for helpful discussions.
This work has been supported by BMBF (OR 06RY9191).

%



%

\end{document}